\documentclass[twoside,twocolumn,9pt]{article}
\usepackage{extsizes}
\usepackage[super,sort&compress,comma]{natbib} 
\usepackage[version=3]{mhchem}
\usepackage[left=1.5cm, right=1.5cm, top=1.785cm, bottom=2.0cm]{geometry}
\usepackage{balance}
\usepackage{mathptmx}
\usepackage{sectsty}
\usepackage{graphicx} 
\usepackage{lastpage}
\usepackage[format=plain,justification=justified,singlelinecheck=false,font={stretch=1.125,small,sf},labelfont=bf,labelsep=space]{caption}
\usepackage{float}
\usepackage{fancyhdr}
\usepackage{fnpos}
\usepackage[english]{babel}
\addto{\captionsenglish}{%
  
}
\usepackage{array}
\usepackage{droidsans}
\usepackage{charter}
\usepackage[T1]{fontenc}
\usepackage[usenames,dvipsnames]{xcolor}
\usepackage{setspace}
\usepackage[compact]{titlesec}
\usepackage{hyperref}

\usepackage{epstopdf}

\definecolor{cream}{RGB}{222,217,201}

\begin{document}

%%%PAGE SETUP%%%
\makeFNbottom
\makeatletter
\renewcommand\LARGE{\@setfontsize\LARGE{15pt}{17}}
\renewcommand\Large{\@setfontsize\Large{12pt}{14}}
\renewcommand\large{\@setfontsize\large{10pt}{12}}
\makeatother

\setcounter{secnumdepth}{5}

\makeatletter 
\renewcommand\@biblabel[1]{#1}            
\renewcommand\@makefntext[1]% 
{\noindent\makebox[0pt][r]{\@thefnmark\,}#1}
\makeatother 
\renewcommand{\figurename}{\small{Fig.}~}
\sectionfont{\sffamily\Large}
\subsectionfont{\normalsize}
\subsubsectionfont{\bf}
\setstretch{1.125} 
 
\setlength{\jot}{10pt}
\titlespacing*{\section}{0pt}{4pt}{4pt}
\titlespacing*{\subsection}{0pt}{15pt}{1pt}

%%%TITLE AND ABSTRACT (CLEANED)%%%
\title{\textbf{Particle-Wall Alignment Interaction and Active Brownian Diffusion Through Narrow Channels$^\dag$}}

\author{%
  Poulami Bag,$^{a}$ Shubhadip Nayak,$^{a}$ and Pulak Kumar Ghosh$^{a\ddag}$ \\[0.5em]
  \small $^{a}$Department of Chemistry, Presidency University, Kolkata - 700073
}
\date{}  

\twocolumn[
  \begin{@twocolumnfalse}
    \maketitle
    \begin{abstract}
    We numerically examine the impacts of particle-wall alignment interactions on active species diffusion through a structureless narrow two-dimensional channel. We consider particle-wall interaction to depend on the self-propulsion velocity direction whereby some specific particle's alignments with respect to the boundary walls are stabilized most. Further, the alignment interaction is meaningful as long as particles are close to the confining boundaries. Unbiased diffusion of active particles for various possible stable velocity alignments against the walls has been examined. We show that for the most stable configuration leading to self-propulsion velocity direction perpendicular to the wall, diffusivity becomes inversely proportional to the square of alignment interaction torque. On the other hand, when the self-propulsion velocity direction making an acute angle to the channel walls is the most stable configuration, diffusion exponentially grows with strengthening alignment interaction. Hence, particle-wall interaction plays a pivotal role in the transport control of active particles through narrow channels. Moreover, the impacts of the alignment interactions on diffusion largely depend on the particle's self-propulsion properties and its chirality. Our simulation results can potentially be used to understand unbiased diffusion of artificial or living micro/nano-objects (such as virus, bacteria, Janus particles, etc.) though narrow confined structures.
    \end{abstract}
    \vspace{0.6cm}
  \end{@twocolumnfalse}
]

%%%FONT SETUP%%%
\renewcommand*\rmdefault{bch}\normalfont\upshape
\rmfamily

%%%FOOTNOTES%%%
 
\footnotetext{\dag~Email: pulak.chem@presiuniv.ac.in}

\section{Introduction}
Diffusion of a tracer (active or passive) in a confined structure attracted widespread interdisciplinary interest~\cite{Hanggi-diffusion,photo-1,photo-2,photo-3,photo-4} for its own merit. For passive particles, confining boundaries affect dynamics through particle-wall energetic interactions~\cite{F-Durst,Yueming-Wang}, entropic effects~\cite{Zwanzig,Hanggi-diffusion,Bosi,Ghosh,Schuss,Holcman1,AiBQ,BAi3,Debasish,Debasish2,Debasish3,Dagdug,Berezhkovskii}, pure geometric effects due to sharply uneven boundaries~\cite{geometricSR1,Makhnovskii,sevali}, as well as, non-trivial particle-wall hydrodynamic interaction~\cite{Marchesoni-hy,Misiunas-hy}. In contrast to the passive particles, active ones have an additional degree of freedom, the direction of self-propulsion velocity. Further, the orientational relaxation timescale of self-propulsion velocity~\cite{{Howse,Romanczuk1,Marchetti,Bechinger,Lowen,Hagen2,Volpe1,Volpe2,Li1,wang3}} dictates dynamics of active particles.
Thus, for this kind of species, boundary effects can enter into dynamics through particle-wall alignment interaction.  It is perceivable that even a little change in the direction of self-propulsion due to alignment interaction near the wall can produce significant changes in diffusion and other transport quantifiers.

In this paper, our attention is restricted to exploring the possible impacts of particle-wall alignment interaction on the unbias transport of active particles through a narrow channel. Active particles are capable of extracting kinetic energy with persistent motion thanks to their built-in functionality. Most popular design for artificial active particle is self-propelled Janus particles~\cite{{review,cataly1,cataly2,cataly3,ther,mag,sano2,Yu1,Wang1,Wang2,zhu}}. Such particles possess two distinct faces (like the Roman god Janus) with different physical/chemical/optical properties. Our investigation aims to understand the transport of both artificial self-propelled particles (e.g., Janus particles) as well as the natural ones through a narrow channel, considering the interaction of their velocity alignment with the channel walls.

Earlier studies~\cite{Caprini,Das1,Fily1,Debnath,MSshort,ratchet,GNM} show that active particles exhibit unusual transport features and non-equilibrium behaviors in confined structures.  Autonomous directed motion in asymmetric periodic channels~\cite{MSshort,ratchet,Lan}, absolute negative mobility ~\cite{GNM}, capture and separation of active particles~\cite{Daisuke,pbag,Misko1}, and quorum sensing active matter in a confined geometry~\cite{Yuxin} are to mention a few where the boundary effects play dictating roles in dynamics. All these studies provide suggestive opinions about transport control in confined structures.    

% \paragraph{This is the next level heading.~~} For this level please use \texttt{\textbackslash paragraph}. These headings should also end in a full point.

% \section*{}
% \subsection{Graphics}
% Graphics should be inserted on the page where they are first mentioned (unless they are equations, which appear in the flow of the text). 

\begin{figure}
\centering
 \includegraphics[width=0.40\textwidth,height=0.2\textwidth]{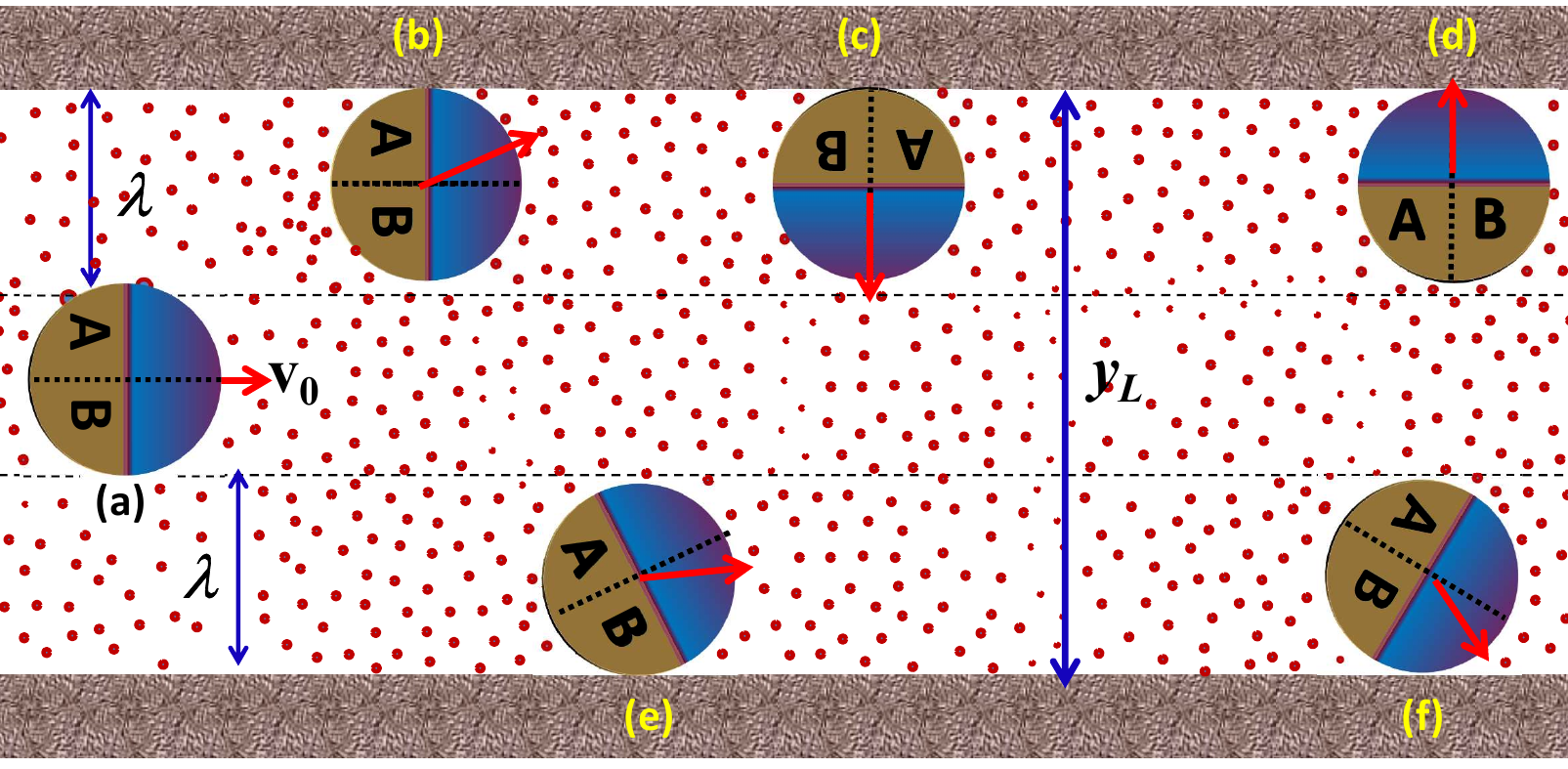}
\caption {(Color online) Conceptual drawing depicting local fuel density around the coated active surface (dark yellow) of a spherical Janus particle for its different configurations in a narrow channel. Janus particles  acquire self-propulsion through self-phoresis processes taking place at the active surface.  To better realize the uneven rate of self-phoresis near the boundary walls, we imagine two symmetric parts $A$ and $B$ of the coated hemisphere of the particle. In the bulk [see (a)], the self-propulsion velocity is pointed along the symmetry axis which is indicated by a dotted line bisecting the Janus particle. The different possible orientations of the Janus particle with respect to the walls are represented in (b)-(f).  For (b), (e), and (f), the parts $A$ and $B$ are not equally exposed to the fuel. Thus, for these alignments self-propulsion velocity orientation may not fall on the particle's symmetry axis (dotted lines) like in (a).  Red arrows show possible directions ${\vec v_0}$ for the all configurations (a-f). Here,  $y_L$ is the channel width and $\lambda$ is the cut-off distance from the boundary below which the particle's dynamics is affected by the wall.
  }
\end{figure}

%%%%%%%%%%%%%%%%%%%%%%%%%%%%%%%%%%%%%%%%
Our study considers diffusion of overdamped active particles through a narrow structureless\cite{note} straight channel having a width about five to ten times larger than the particle diameters. We assume that when particles get close to the walls, the self-propulsion direction is affected due to particle-wall interactions. In case of active particles of Janus kind, intuitively, the alignment effect can be considered as a result of the following facts:   (i) The materials at two hemispheres of Janus particles possess different dielectric properties, thus, it is expected that particle-wall interactions may stabilize more at some specific orientations. (ii) Near the confining boundary, the entire coated surface may not be exposed to the fuel equally (see Fig.1). This may produce an unequal rate of self-phoretic process at the parts $A$ and $B$ of the coated hemisphere (see Fig.1). Thus, direction of self-propulsion velocity can be affected as the active particles approach to the wall.  Figure 1 illustrates the plausible direction of self-propelled velocity $\vec{v_0}$ (at different configurations) due to the uneven rate of self-phoretic processes over the active surface.  When the JP is away from the wall [Fig.1(a)] the rate of self-phoretic process in the part $A$ and $B$, becomes equal. This results in self-propulsion velocity to be directed along a symmetry axis as indicated by the red arrow. However, for the configurations shown in Fig.1(b, e, and f), it is apparent that the fuel density around parts $A$ and $B$ of the coated hemisphere are not the same. Thus, for these configurations, $\vec{v_0}$ direction does coincide on the certain symmetry axis of the particle [see supporting information (see SI-6)].
%%%%%%%%%%%%%%%%%%%%%%%%%%%%%%%%%%%%%%%%
\begin{figure}
\centering
\includegraphics[width=0.44\textwidth,height=0.33\textwidth]{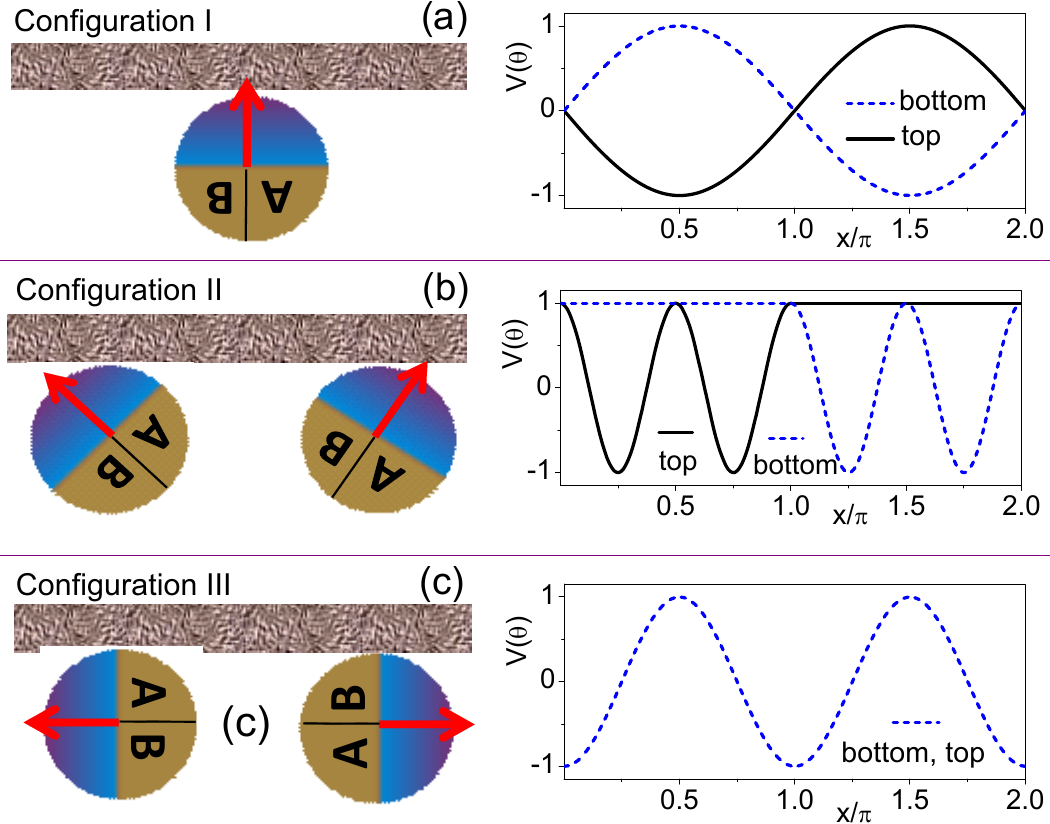}
\caption {(Color online) Schematic illustration of  different stable configurations and associated $\vec{v_0}$ orientation dependent interaction potential.  (a) {\it Configuration I}: Interaction potential is given by Eq.~(\ref{wall1}) with $\tilde{d}=1$ and $\phi = 0$. (b) {\it Configuration II}: Interaction potential is represented by Eq.~(\ref{potentail45t}-\ref{potentail45b}). (c) For {\it Configuration III}, interaction potential is given by Eq.~(\ref{wall1}) with $\tilde{d}=2$ and $\phi = \pi/2$. }
\end{figure}

The microscopic details of the self-propulsion mechanism are explored in Ref.~\cite{Bechinger,Hagen2,Jeffrey, Pierre, Ehud, Arango}. It is assumed that the self-propulsion velocity is generated from the ``effective'' force~\cite{Hagen2,Bechinger} exerted by the suspension fluid on the coated active surface of the particle at low Reynolds numbers. Unlike the case of externally applied force, the center of self-propulsion force does not necessarily coincide with the particle's center of mass. It is more likely to happen near the wall due to the uneven rate of the self-phoretic processes over the active surface.  Thus, the propulsion force may tend to rotate the overdamped active particle around its center of mass~\cite{eccentric1,eccentric2}. 

To illustrate the plausible origin of the wall-particle alignment interactions (in Fig.1 and Fig.2), we consider spherical Janus-kind active particles that appear achiral. However, the alignment interaction is expected to occur even for the active particles with chiraily. Previous studies \cite{Uspal,Bianchi,Mozaffari,Poddar,Czajka} have shown that both chiral and achiral active particles tend to prefer specific orientations with respect to the walls. The alignment interaction with the walls arises from the interplay between the energetic interaction of the particles with the walls and hydrodynamic effects \cite{Uspal,Bianchi,Kreissl1,Kreissl2}, regardless of whether the particles are achiral or not. Furthermore, chirality plays an important role in biological and chemical processes, and it can trigger phenomena\cite{Kraft,Kummel,Ambarish,Schamel,Collins} that are otherwise difficult to observe. Therefore, to generalize our analysis, we are considering the impact of particle-wall alignment interaction on diffusion through narrow channels for both chiral and achiral active particles.

Following the above discussion, a simplified model has been considered to take into account the different behaviors of particles' self-propulsion velocity direction near walls. %To be specific, we consider particle-wall alignment interaction whereby self-propulsion gets stabilized at some specific orientations.
 For quantitative  analysis, our study encodes the following forms of alignment interactions concerning the top ($V_{t}$)  wall, 
\begin{eqnarray} 
V_t(\theta)=-\tilde{\omega}(y) \sin{(\tilde{d}\;\theta+\phi)}.
\end{eqnarray}\label{wall1}
Where, $\theta$ is the direction of $\vec{v_0}$ with respect to the channel axis and $\tilde{\omega}(y)$ represents the strength of alignment interaction. Here, $\tilde{d}$ is the degeneracy factor, which determines the number of equivalent orientations that appear for rotation over the angle $2\pi$. The influence of alignment interaction is limited to a certain distance from the wall. The most stable orientations for $\vec{v_0}$ near walls are determined by the phase factor $\phi$ along with $\tilde{d}$. The alignment interaction potential against the bottom wall, $V_{b}(\theta)$, has a similar orientation dependence as $V_{t}(\theta)$.  
%For instances, when $\tilde{d} =1$,  $\phi=0, \pi/4, \; {\rm and}\; \pi/2 $,  respectively correspond to the most stable orientations of $\vec{v_0}$ near the upper boundary at $\theta = \pi/2, 3\pi/4, \; {\rm and} \; \pi$.  

Among the very many possible stable $\vec{v_0}$ orientations near the wall, we mainly focus on the following three cases. Schematics and associated interaction potentials for these configurations are shown in Fig. 2.
\newline (A) { \it Configuration I}:  It considers the stable orientation of self-propulsion velocity perpendicular to the wall. The alignment interaction potentials with the top wall is presented by Eq.~(\ref{wall1}), with $\phi=0$ and $\tilde{d} = 1$. Here, the interaction energy with the bottom wall, $V_{b}(\theta)$,  is given by, $V_{b}(\theta)=-V_{t}(\theta)$.   
\newline (B) In { \it Configuration II}, it is assumed that the most stable alignment of $\vec{v_0}$ against the top wall is at angles of $\theta = \pi/4$ and $3\pi/4$ [see Fig. 2(b)]. % Note that due to structural symmetry of Janus particles,  both the orientations, $\theta = \pi/4$ and $3\pi/4$ have the same alignment interaction energy.
  However, near the bottom wall, the active particles get stabilized most at two equivalent orientations, $\theta = -\pi/4$ and $-3\pi/4$. In reference to Eq.~(\ref{wall1}), the associated alignment interaction potential for this configuration is four-fold degenerate, with $\tilde{d} = 4$ and a phase factor of $\phi = 3\pi/2$. 
 Details about the potential presented in Sec.~IIIB. 
\newline (C)  {\it Configuration III}: Here the most stable self-propulsion velocity alignments near the walls occur at the two equivalent orientations, $\theta = 0$ and $\pi$. Associated alignment interaction potential (\ref{wall1}) has a two-fold degeneracy ($\tilde{d} = 2$)  with $\phi = \pi/2$ and $V_{b}(\theta)=V_{t}(\theta)$.  

In the stable configurations where the self-propulsion velocity is directed away from the walls (e.g., $-\pi < \theta <0$ and $ 0< \theta <\pi $, with respect to the top and bottom walls, respectively), active particles tend to move away from the regions of alignment interaction quickly.  As a result, the alignment interactions have no noticeable impact on the dynamics of the active particles. Therefore, we mainly focus on the three stable configurations mentioned above.  Further note that when ${\vec v_0}$ making an acute angle against the walls is the most stable configuration, active particles display similar diffusion features as in {\it Configuration II}.  

%We numerically explore diffusion mechanisms for the three types of stable ${\vec v_0}$ orientations: Configuration I, II and III. 
 Our simulation results indicate that when the strength of the alignment interaction is comparable to, or slightly greater than the rotational diffusion, the diffusion behavior of active particles undergoes significant changes.  For the stable Configuration I, diffusion follows an inverse relation with the square of alignment interaction strength.
On the other hand, when stable $\vec{v_0}$ orientation corresponds to Configuration II and III,  unbiased transport featured with very long diffusion transients and diffusivity exponentially grows with the alignment interaction strength up to a certain threshold value. Further strengthening interaction beyond the threshold value, diffusion gradually approaches an asymptotic limit. Additionally, the diffusion features of active particles largely depend on their chirality, unless the alignment interaction torque is too strong in comparison to the chiral torque.

 %Further, we have demonstrated that in presence of wall-particle alignment interaction, particles' apparent weight or little top-bottom asymmetry in alignment interaction suffice to rectify motion chiral swimmers. Conspicuously, such autonomous directed motion does not require any spatial periodic structure.     

This paper is organized as follows. In Sec.~II, we describe the model and discuss the significance of its relevant parameters. We then present our key numerical results and provide some analytic arguments in Sec.~III, where we also perform a detailed analysis of the diffusion mechanisms for Configuration I, II, and III in sub-sections IIIA, IIIB, and IIIC, respectively. Finally, in Sec.~IV, we summarize our results and provide some concluding remarks.

%Chirality\cite{Bianchi,Uspal,Mozaffari,Poddar,Kreissl1,Kreissl2,Collins, %Ambarish,Schamel} test%%5\cite{Ambarish,Collins,Kreissl1,Kreissl2,Czajka,Uspal,Mozaffari,Poddar,Schamel} 
 
\section{Model}
Let us consider an active Brownian particle diffusing in a two-dimensional (2D) narrow straight channel in the absence of any external biases. To avoid unessential complications, we restrict our analysis to the case of 2D channel.  We anticipate that some of the findings can be easily extended to 3D channels with flat walls. However, for curved surfaces, the interaction between particle-wall alignment interaction would be complex, and the results in 2D could be significantly different from the 3D cases.

 In the channel, particle's dynamics is governed by boundary effects in addition to the self-propulsion and thermal fluctuations.  We encoded dynamics of the particle's center of mass ($x,y$) by the following set of overdamped Langevin equations, 
\begin{eqnarray}
\dot{x}&=&v_0 \cos{\theta}+\sqrt{2D_0}\;\xi_{x}(t)\label{L1} \\
\dot{y}&=&v_0\sin{\theta}+\sqrt{2D_0}\;\xi_{y}(t)\label{L2}\\
\dot{\theta} &=& \Omega + \omega(y)\cos(\tilde{d}\;\theta+\phi)+\sqrt{2D_\theta}\;\xi_\theta(t)\label{L3}.
\end{eqnarray}
The self-propulsion velocity with a constant modulus  $v_0$ is oriented at an angle $\theta$ with respect to the channel axis (x-axis). Where, $\theta$ evolves in time according to Eq.~(\ref{L3}). The last terms in Eq.~(\ref{L1}-\ref{L2}), $\xi_{x}(t)$ and $\xi_{y}(t)$, are thermal noises with Gaussian distribution. Their other statistical properties are characterized as follows, 
\begin{eqnarray}
\langle \xi_{x}(t) \rangle &=&\langle \xi_{y}(t) \rangle = 0, \\ \nonumber 
\langle \xi_{x}(t) \xi_{x}(t') \rangle &=& \langle \xi_{y}(t) \xi_{y}(t') \rangle  = \delta({t-t'}).\nonumber 
\end{eqnarray}
 The strength of these thermal fluctuations, $D_0$ is a measure of translational diffusion in the bulk in the limit $v_0 \rightarrow 0$. For a spherical particle with radius $r_0$ , $D_0$ is related to temperature ($T$) and medium viscosity ($\eta$) as $D_0 = k_B T/6\pi\eta r_0$. Where, $k_B$ is the Boltzmann constant.  The noise term $\xi_\theta(t)$ in the Eq.~(\ref{L3}) is responsible for rotational diffusion follows a similar statistical as thermal translational noises [$\xi_{x}(t)$ and $\xi_{y}(t)$]. Rotational diffusion of a colloidal particle in a free space can be estimated based on Einstein--Smoluchowski relation, $D_{\theta}=k_B T/\gamma_R$, with $\gamma_R = 8\pi\eta r_0^3$.  However, for an active particle, the diffusive rotational motion may depend on the self-propulsion mechanisms. Thus, our study considers $D_\theta$ as an independent model parameter.

 The rotational dynamics of the active particle is governed by intrinsic chiral torque $(\tilde{\Omega})$ and alignment interaction near the wall in addition to the fluctuation-induced diffusion. In the supporting information (SI-1), we discuss about chirality of active particles.  The alignment interaction-induced torque is derived from the orientation-dependent potential Eq.(\ref{wall1}).  The amplitude of this torque,  $\tilde{\omega}(y)$,  depends on the wall-particle separation. In the overdamped limit angular velocity associated to the chirality and alignment interactions are $\Omega=\tilde{\Omega}/\gamma_R$ and  $\omega(y) = \tilde{\omega}(y) /\gamma_R$, respectively\cite{note2}.   We assume the alignment interaction becomes operational as soon as the distance from the wall is less than a cut-off distance $\lambda$.  Moreover, the strength of this interaction decreases exponentially as the particles move away from the walls, 
\begin{eqnarray}
 \omega(y)&=&\omega_0 \exp[-\chi |d(y)-\sigma/2|], \; \; \; {\rm if} \; \; d(y) \le \lambda. \\
 &=& 0 \; \; \; \;   {\rm otherwise} \nonumber
\end{eqnarray}
 Here, $1/\chi$ has similar significance as Debye length, and $\sigma$ is the particle's diameter. The alignment interaction strength assumes its maximum value $\omega_0$ at the separating distance, $ d(y)=\sigma/2$.

 As the direction of the self-propulsion fluctuates in time, its components, $v_0 \cos\theta(t)$ and  $v_0 \sin\theta(t)$ can be considered as color noises with non-Gaussian distribution. In the  free space, the correlation of the self-propulsion velocity components can be expressed as, 
\begin{eqnarray}
\langle \cos\theta(t)\cos\theta(0) \rangle =  \langle \sin\theta(t)\sin\theta(0) \rangle  
= \frac{1}{2}\cos(\Omega t)e^{-D_\theta t}
\end{eqnarray}
Thus,  $1/D_\theta$ can be assumed as rotational relaxation time ($\tau_\theta$) of an achiral particle with the self-propulsion length  $l_\theta = v_0 \tau_\theta$.

 To get particle's position $\{x(t), y(t)\}$ as a function of time, the coupled differential equations (\ref{L1}-\ref{L3}) have been numerically integrated using a standard Milstein algorithm \cite{Kloeden}. To ensure numerical stability, a very short integration time step, $10^{-3} -10^{-4}$, has been used. At the start, t = 0, the particle's self-propulsion velocity orientation is assumed to be uniformly distributed over the range $ 0 \; {\rm to} \; 2\pi$. Each trajectory is allowed to evolve over the time $10^3 \times 1/D_\theta, \; {\rm or} \; 10^3 \times 1/\Omega_0, \; {\rm or} \; 10^3 \times 1/\omega_0, \; {\rm or} \; 10^5$ whichever is greater so as one can safely discard the effects due to transients. All the results reported in this paper have been obtained by ensemble averaging over $10^3 - 10^5$ trajectories depending upon the values of parameters. We provide simulation details in the supporting information SI-4.

We numerically estimate diffusion coefficient ($D$) of active particles along the channel axis.  This transport quantifier is defined as, 
\begin{eqnarray}
D = \lim_{t \rightarrow\infty} \frac{\langle\left[x(t)-x(0)\right]^2\rangle}{2t}
\end{eqnarray}
Here, $\langle ...\rangle$ indicates ensemble averaging. In free space with a constant modulus of self-propulsion velocity, the mean square displacement of a chiral active particles is given by,
\begin{eqnarray} \label{furth} 
&& \langle \Delta x(t)^2\rangle = 2D_0 t - 2v_0^2\frac{D_\theta \Omega}{(D_\theta^2+\Omega^2)^2}e^{-D_\theta t} \sin\Omega t +  \nonumber 
\\
&& v_0^2\left[\frac{D_\theta t}{D_\theta^2+\Omega^2}+\frac{D_\theta^2-\Omega^2}{(D_\theta^2+\Omega^2)^2}\left(e^{-D_\theta t } \cos\Omega t   -1 \right)\right].
\end{eqnarray}
Thus, the bulk diffusivity of a chiral active particle becomes,
\begin{eqnarray}\label{free-duffusion}
D_{s} = D_0  + \frac{v_0^2 D_\theta}{2(D_\theta^2+\Omega^2)}
\end{eqnarray}
In Sec. III, we use this expression for bulk diffusion as a benchmark to assess the impact of wall-particle alignment interaction on transport characteristics.  

%%%%%%%%%%%%%%%%%%%%%%%%%%%%%%%%%%%%%%%%%%%
\begin{figure}
\centering
\includegraphics[width=0.75\linewidth]{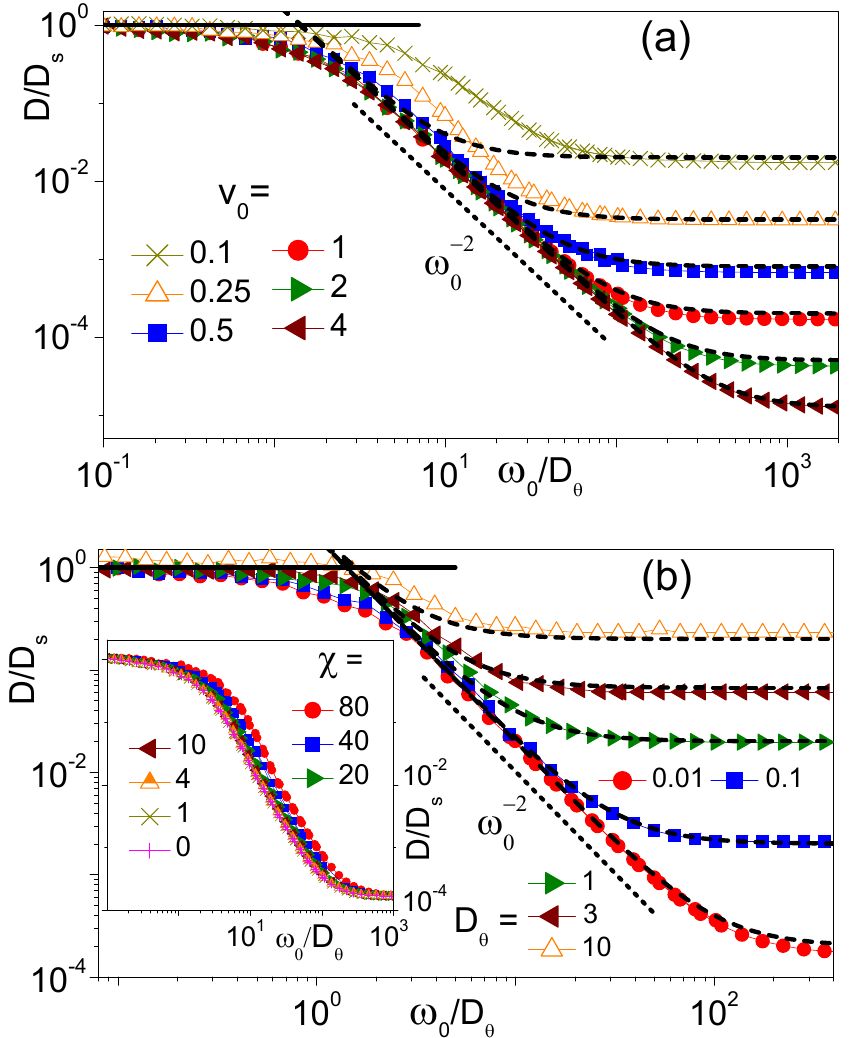}
\caption {(Color online) (a) $D$ versus  $\omega_0$ for different values of self-propulsion velocity $v_0$ (see legends).   (b) Similar plots as panel (a) for $v_0 =1$ and different $D_\theta$ (see legends).  The black dashed lines are analytic estimations based on the Eq.~(\ref{diffusion-pi-by-2}). The solid horizontal line represent diffusion in free space, $D=D_s$ [Eq.~(\ref{free-duffusion}) with $\Omega =0$], and the dotted line is a guide to the eye
showing decay of diffusion constant with the power law, $\omega_0^{-2}$.  Inset here represents $D$ versus  $\omega_0$ for different values of Debye length $1/\chi$.  Simulation parameters (unless reported otherwise in the legends): $v_0 = 1.0, \; D_0 = 0.01, \; \lambda = 0.05,\; \chi = 0, \; \phi = 0, \;\Omega = 0, \; D_\theta = 0.01, \; {\rm and} \; y_L=1$. }
\end{figure}

The diffusivity is estimated for different swimming properties of active particles and particle-wall alignment interactions. The self-propulsion parameters used in our simulations can be experimentally accessible. It becomes apparent by considering times in seconds and lengths in micrometers~\cite{Volpe1,Volpe2}. Our study considers variation of the intrinsic torque $\Omega$ and alignment $\omega_0$ torque over the range $ \{\Omega,\omega_0 \} \ll D_\theta$ to $ \{\Omega,\omega_0 \} \gg D_\theta$. The variation of the channel width is made concerning the self-propulsion length ($l_\theta = v_0/D_\theta$)  and radius of the curvature ($R_\Omega = v_0/\Omega$) for chiral particles. Details about the relevance of the model parameters used in our simulation are discussed in the supporting information SI -2. 

\section{Diffusion}
In Fig. (3-6), we present simulation results that capture the key diffusion features of both chiral and achiral particles through a narrow straight channel. Our results indicate that transport through channels is greatly influenced by the interaction between the particles and the channel walls.  Depending on the direction of stable alignment, diffusion sharply enhances or decays as soon as the torque induced by the alignment interaction surpasses rotational diffusion or the particle's intrinsic torque. We systematically analyze diffusion for three different stable particle-wall alignments: Configuration I,  II, and  III.

\subsection{{ Configuration I:} Stable self-propulsion velocity direction perpendicular to the walls}
Recall that for Configuration I, alignment interaction torque is derived from Eq.~(\ref{wall1}) using  $\phi = 0$ and $\tilde{d}=1$.  Here, $\vec{v_0}$ orientations against the top and bottom walls are most stable at the angles $\theta = \pi/2$ and $-\pi/2$, respectively. For this configuration, Fig. 2 and Fig.3, respectively, depict variation of diffusion constant as a function of alignment interaction strength for achiral and chiral active particles. To begin with we examine the impact of the model parameter  $\chi$ on diffusion. Note that $\chi$ determines how fast the alignment interaction gets blurred as moving away from the boundary walls.  Our simulation results [ see inset of Fig.~3(b)] show that a particle's transport is sensitive to this parameter  as long as  $1/\chi$ is shorter than the cut-off length $\lambda$. However, $D$ versus $\omega_0$ is not affected much by the variation of $\chi$. Therefore, to reduce parameter space, we set $\chi=0$, unless mentioned otherwise.
%%%%%%%%%%%%%%%%%%%%%%%%%%%%%%%%%%%%%%
\begin{figure}[h!] 
\centering
\includegraphics[width=0.75\linewidth]{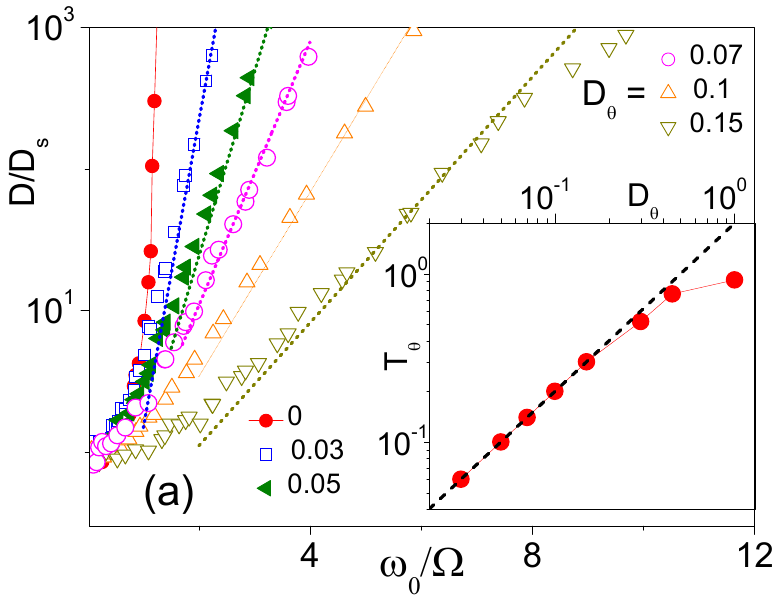}
\includegraphics[width=0.75\linewidth]{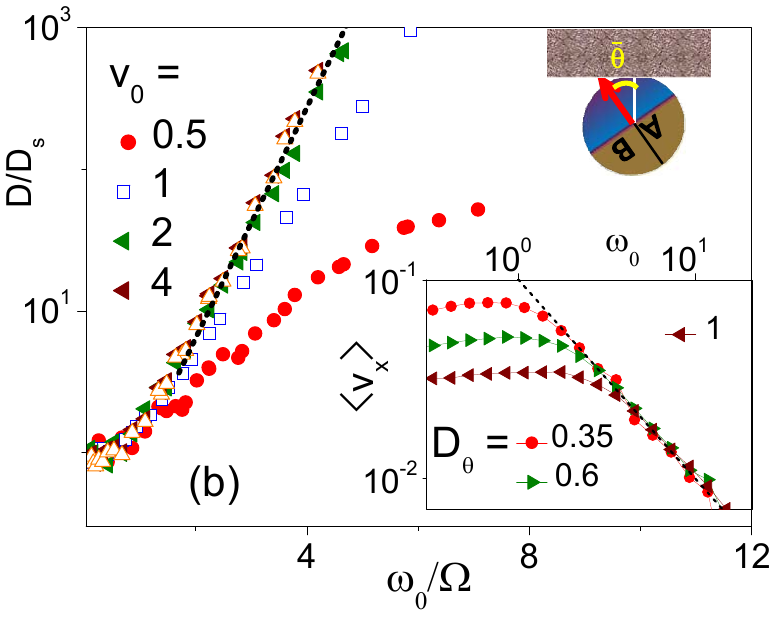}
\includegraphics[width=0.75\linewidth]{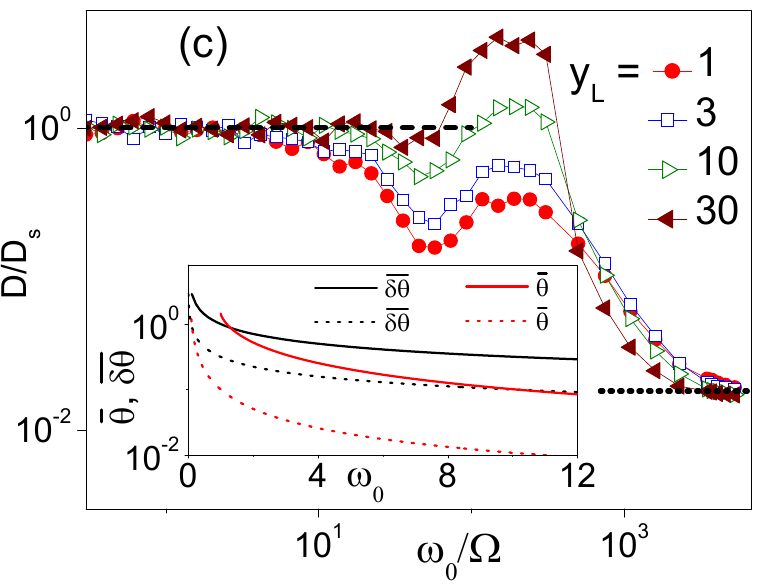}
\caption {(Color online) (a) Diffusion constant of a chiral active particle as a function of $\omega_{0}/\Omega$  for different $D_\theta$ (see legends) and $\Omega = 0.3$. The exponential growing branch fitted with $\exp(\omega_0/2T_\theta)$ [dotted lines] to extract value of $T_\theta$. The inset shows variation of $T_\theta$ with $D_\theta$. The dashed line depicts $T_\theta = D_\theta$. (b) Similar plots as the panel (a) but for different $v_0$ (shown in the legends) and $D_\theta$ is fixed at $0.1$. Here, the factors $T_\theta$ in the exponential fitting is  insensitive to $v_0$.   The inset depicts variation of $\langle v_x \rangle $ near the top-wall. The decaying tail is fitted with the power law $\omega_0^{-1}$. (c) Similar plots as the panel (a) but for different channel width (see legends) and $D_\theta =1$. Diffusion in the asymptotic limits $\omega_0/D_\theta \gg 1$ and $\omega_0/D_\theta \ll 1$ are respectively marked with dotted and dashed lines 
[using Eq.~(\ref{free-duffusion}) and Eq.~(\ref{diffusion-pi-by-2a})]. The inset compares variation of the average title angle $\overline{\theta} $  [$\Omega = 1$ (solid line), $0.1$ (dotted line)]  and standard deviation of orientation fluctuation $\overline{\delta \theta}$ [$D_\theta = 1$ (solid line), $0.1$ (dotted line)] . Simulation parameters (unless reported otherwise in the legends): $v_0 = 1.0, \; D_0 = 0.01, \; \lambda = 0.05,\; \chi = 0, \; \phi = 0, \;\Omega = 0.3, \; D_\theta = 1.0, \; {\rm and} \; y_L=1$. }
\end{figure}

\subsubsection{Diffusion of achiral active particle} Inspecting simulation results presented in Fig.~3(a,b), we note the key diffusion features of an achiral active particle. (i) As long as $\omega_0 \ll D_\theta$, the ratio,  $D/D_s \approx 1$ [indicated by solid horizontal line in Fig.~3(a,b)].  This suggests that in the weak alignment coupling region, an achiral active particle diffuses through the channel with the free space diffusivity, $D \sim D_0 + v_0^2/2 D_\theta$ [Eq.(\ref{free-duffusion}) with $\Omega = 0$]. (ii) When the strength of the alignment interaction becomes stronger than the rotational diffusion, the diffusivity decreases following a power law of $\omega_0^{-2}$ [as shown by the dotted lines in Fig.~3(a,b)]. (iii) As $\omega_0$ approaches to infinity, diffusion becomes insensitive to the alignment interaction and reaches a constant value.

In the strong alignment coupling limit, the diffusion features (ii) and (iii) can be understood as follows.  When $\omega_0 \gg D_\theta$ and $v_0$ much stronger than thermal fluctuations $D_0$, the active particle assume the stable orientation where $\vec{v_0}$ is perpendicular to the walls. As a result, the self-propulsion force pushes the particle against the walls [see Fig.~2(a)]. Under this situation, diffusion can occur along the channel walls due to fluctuations of $\vec{v_0}$ orientation around the stable angles $\theta = \pm \pi/2$, as well as thermal translational noises.  The probability of a small $\vec{v_0}$ direction fluctuation, $\delta \theta $, about the most stable state is given by, $p(\delta \theta) = 1/\sqrt{2\pi D_\theta/\omega_0} \exp(-\omega_0 \delta \theta^2/2D_\theta)$. Note that this expression of distribution is obtained assuming linearization of potential $V_{t/b}$ around the minimum. Where, $\omega_0 = V_{t}''(\pi/2)=V_{b}''(-\pi/2)$ and mean square deviation about the stable orientation , $\langle \delta \theta^2 \rangle = D_\theta/\omega_0$ [see supporting information (SI-5)]. When the particles are stuck on the top wall, it can be assumed that, on average, the self-propulsion velocity switches between the angles $\pi/2 \pm \sqrt{D_\theta/\omega_0} $ over an approximate period $1/\omega_0$. The active particle appears to randomly switches between kinematic states [illustrated in Figure S2(c) in the supporting information (SI-5)] with opposite effective x-directional velocity, $\pm v_0 \sin\sqrt{D_\theta/\omega_0}$ and relaxation time $1/\omega_0$. The diffusion constant for such a process is given by  \cite{gardiner},
\begin{eqnarray}
D = v_0^2 \sin^2\left(\sqrt\frac{D_\theta}{\omega_0}\right) \; \frac{1}{\omega_0} + D_0
\end{eqnarray}
The last term here accounts for thermal translational diffusion. When $D_\theta \ll \omega_0$, $\sin(\sqrt{D_\theta/\omega_0}) \sim \sqrt{D_\theta/\omega_0}$, thus the above equation is simplified as,
\begin{eqnarray}\label{diffusion-pi-by-2}
D = \frac{v_0^2  D_\theta}{\omega_0^2} + D_0
\end{eqnarray}
Further, for $\omega_0 \rightarrow \infty$, the self-propulsion velocity direction is locked so strongly that its component along the channel axis due to orientational fluctuations becomes vanishingly small. As a result, the diffusion occurs due to thermal translational motion only,
 \begin{eqnarray}\label{diffusion-pi-by-2a}
D  \sim  D_0
\end{eqnarray}
These estimations accord well with numerical simulation results (presented in Fig.3) when $v_0 \gg D_0$ and $\omega_0 \gg D_\theta$.

\subsubsection{Diffusion of a chiral active particle}
It is apparent from Fig.4 that chiral active particle's diffusion is notably different from achiral ones.  Further, diffusion behavior for $\Omega > D_\theta$ noticeably differs from its opposite limit, $\Omega < D_\theta$. When the intrinsic chiral torque $\Omega$ is stronger than $D_\theta$, the diffusion through the channel grows exponentially with strengthening particle-wall alignment interaction. The exponent depends on the $\Omega$ and $D_\theta$, however, insensitive to the self-propulsion velocity as long as $v_0 \gg D_0$ [see Fig.~4(a-b)]. On the other hand, when rotational diffusion is much stronger than $\Omega$,  $D \; vs.\;\omega_0$  passes through a minimum [see Fig.~4(c)]. For both of these regimes of parameter, diffusion decays to the asymptotic value, $D \sim D_0$ when $\omega_0 \gg \{D_\theta, \; \Omega\}$.  These intriguing diffusion features of chiral active particles can be justified based on the following considerations.

Recall that when $\omega_0$ is much weaker than either of $D_\theta$ or $\Omega$, diffusion through the channel can be estimated using Eq.~(\ref{free-duffusion}). For $\Omega > D_\theta $, when $\omega_0$ surpasses intrinsic torque, particles tend to get aligned against the wall as soon as they approach the boundaries. At the stable ${\vec v_0}$ orientation, the alignment interaction torque is counterbalanced by the intrinsic torque, i.e.,  $ \cos\theta = -\Omega/\omega_0$. Thus, the intrinsic chiral torque makes the self-propulsion velocity direction tilted by an angle, $\overline{\theta} = -\sin^{-1}\left(\Omega/\omega_0\right)$ with respect to the normal to the channel wall [see sketch in the inset of Fig.4(b)]. Due to the tilting of ${\vec v_0}$ orientation, chiral active particles keep sliding along the channel walls with the self-propulsion velocity component $-v_0 \sin{\overline{\theta}}$. On the top (bottom) wall, a levogyre active particle tends to slide along the positive (negative) direction. Such sliding leads to a very long diffusion transient. Transition from the stable sliding states (i.e., reversal of self-propulsion velocity direction) requires uphill rotational diffusion by an appropriate angle or thermal diffusion over the length $\lambda$ against the self-propulsion force. Assuming Arrhenius type barrier crossing process from the stable configuration, the direction reversal time can be expressed as,   
\begin{eqnarray}\label{tauR}
\tau_R = A \exp\left(\omega_0/2D_\theta \right),
\end{eqnarray}
where, A is insensitive to the rotational diffusion, however, it depends on the potential energy curvature about the stable configuration. Estimation in Eq.~(\ref{tauR}) ignores the change of barrier height due to $\Omega$. It requires alignment interaction induced torque is much stronger than $\Omega$.
Thus, the diffusivity of a chiral active particle can be approximated as~\cite{gardiner}, $D \sim B v_0^2 \tau_R$. The dimensionless factor $B$ weakly depends on $\omega_0$ and $\Omega$. 
This estimation well justifies the exponential growth of $D$ with $\omega_0$.
Further, based on the least square fitting of numerical diffusion data we extract the best-fitted exponents in $D$ versus $\omega_0$, which is close to $1/2D_\theta$ [see inset of Fig.4(a)], as it is predicted by the Eq.~(\ref{tauR}).

For $ D_\theta > \Omega $, standard deviation of orientation about the stable configuration, $\overline{\delta \theta} \sim \sqrt{D_\theta/\omega_0}$ is larger than the tilting angle $\overline{\theta}$. The variations of $\overline{\delta \theta}$ and $\overline{\theta}$ with $\omega_0$ have been compared in the inset of Fig.4(c).  When $\omega_0$ is not too large, a very small rotational fluctuation suffices to flip the self-propulsion direction about the stable angle.  As a result,  as soon as the  particles start getting aligned against the wall, diffusion decreases with increasing $\omega_0$ like achiral particles. However, when alignment interactions get stronger than the rotational fluctuations, particles tend to settle on the wall with a $\vec{v_0}$ direction $\pi/2 \pm \overline{\theta}$. Such orientation makes the particles slide on the wall in a particular direction for quite a long time. This effect enhances diffusion exponentially. Thus, for $D_\theta > \Omega$, $D \; vs. \; \omega_0$ passes through a minimum.

For $\omega_0 \gg \Omega$,  the tilting angle becomes vanishingly small, and the active particle's self-propulsion velocity direction tends to be perpendicular to the channel walls.  The components of self-propulsion velocity (near walls) along the channel axis decay with $\omega_0^{-1}$. It has been depicted in the inset of Fig.4(b). In this limit, for both the situations, $\Omega > D_\theta $ and $\Omega < D_\theta $, the diffusion contribution of self-propulsion motion gets suppressed to zero. Particles exhibit random motion only for thermal translational fluctuations. Thus, diffusivity reaches the limiting value $D_0$.

We conclude this section with remarks about the impact of thermal fluctuations  on particle-wall alignment interaction. The thermal noise-induced diffusion can take particles away from the wall. However, it is a threshold crossing event with a barrier $\sim v_0 \lambda$. Thus, thermal fluctuations noticeably reduce the impact of wall-particles interaction in the dynamics as soon as $v_0 \lambda$ becomes comparable to $D_0$.

%\subsection{Configuration II: Stable self-propulsion velocity orientations against top walls at $\theta = \pi/4$ and $ 3\pi/4$ }
\begin{figure} 
\centering
\includegraphics[width=0.75\linewidth]{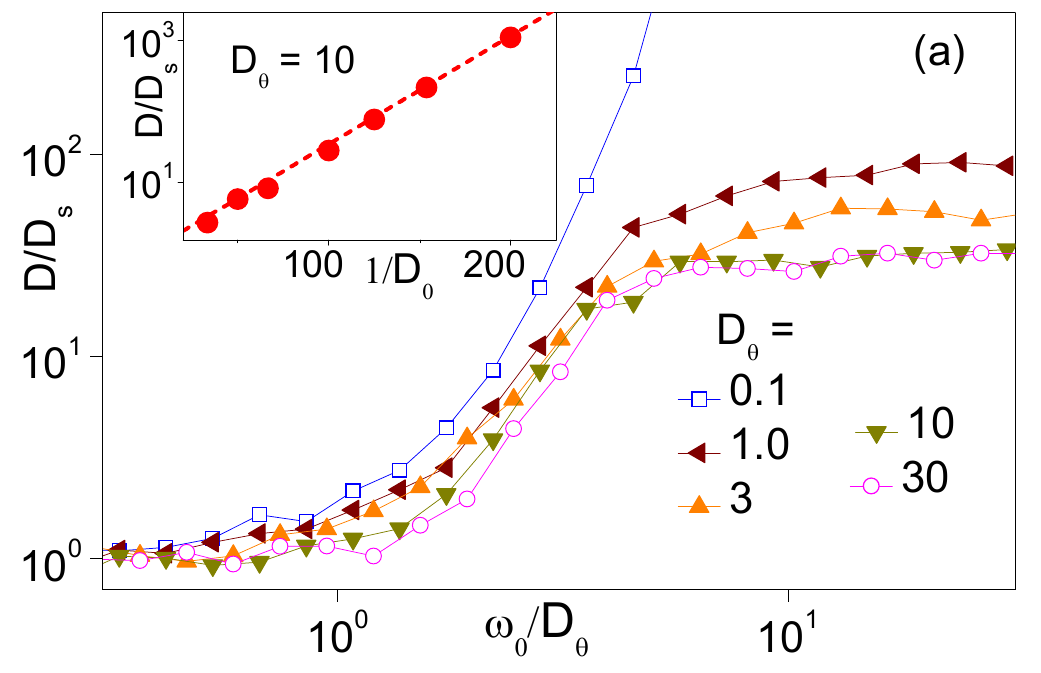}
\includegraphics[width=0.75\linewidth]{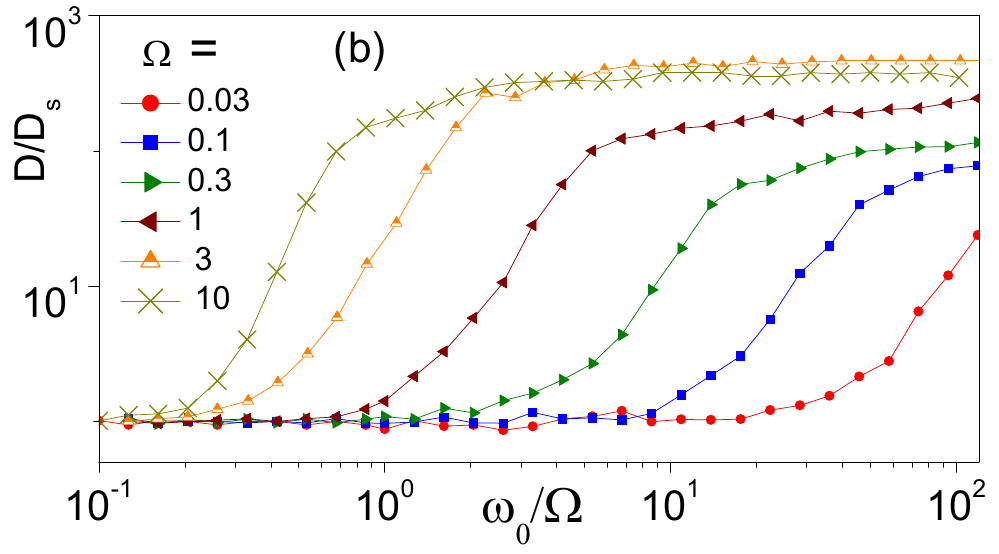}
\caption {(Color online) (a) $D \; vs. \; \omega_0$ of achiral active particles for the Configuration II. Inset: $D_\infty$ (diffusion in the limit $\omega_0 \rightarrow \infty$) $vs.\; 1/D_0$ for $y_L \gg l_\theta$.   (b) $D \; vs. \; \omega_0$ for different chiral torque $\Omega$ (see legends) and $D_\theta = 1$.  Simulation parameters (unless reported otherwise in the legends): $v_0 = 1.0, \; D_0 = 0.01, \; \lambda = 0.05,\; \chi = 0, \;\Omega = 0, \; D_\theta = 0.3, \; {\rm and} \; y_L=1$. }
\end{figure}
\subsection{Configuration II: Stable self-propulsion velocity direction make acute angles against walls}
As a case of stable velocity directions at acute angles, here we consider particle's ${\vec v_0}$  to be most stabilized against the top (bottom) walls when $\theta = \pi/4$ and $ 3\pi/4$ ($\theta = -\pi/4$ and $ -3\pi/4$) [see Fig.~2(b)]. 
%As a case of stable velocity directions at acute angles, here we numerically examine the impacts of wall-particle alignment interaction on diffusion when the most stable configuration against the top (bottom) walls at the self-propulsion velocity orientations $\theta = \pi/4$ and $ 3\pi/4$ ($\theta = -\pi/4$ and $ -3\pi/4$) [see Fig.~2(b)].
 For numerical simulations, we consider the particle-wall alignment interaction potential as follows:
\begin{eqnarray} \label{potentail45t}
V_t(\theta) &=&  \omega(y) \cos(4\theta) \;\; {\rm for} \;\; 0 \leq \theta \leq \pi \nonumber
\\ 
&=& \omega(y) \;\; {\rm otherwise}
\end{eqnarray}
for the top wall and 
\begin{eqnarray} \label{potentail45b}
V_b(\theta) &=& \omega(y) \cos(4\theta) \;\; {\rm for} \;\; \pi \leq \theta \leq 2\pi \nonumber
\\ 
&=& \omega(y) \;\; {\rm otherwise}
\end{eqnarray}
for the  bottom wall. The right panel of Fig.~2(b) depicts $V_t(\theta)$ (solid line) and $V_b(\theta)$ (dotted line).

The plots shown in Fig. 5(a) illustrates the diffusion of achiral active particles as a function of $\omega_0$ for the particle-wall alignment interaction potential~(\ref{potentail45t}-\ref{potentail45b}). Initially, the diffusion coefficient $D$ increases slowly with the increase in alignment interaction strength. However, it eventually exponentially increases and reaches an asymptotic value $D_\infty$ when $\omega_0 >> D_\theta$. The simulation results indicate that $D_\infty$ depends on both $D_0$ and $D_\theta$. Nonetheless, when $l_{\theta} \ll y_L$, $D_\infty/D_s$ becomes insensitive to the rotational diffusion.
 The following facts are associated with these intriguing diffusion features: 
\newline (i) When $\omega_0$ is much larger than $D_\theta$, the direction of the self-propulsion velocity becomes locked at an angle of $\pi/4$ with respect to the walls. Consequently, the active particle continues to slide along the walls until the direction of $\vec{v_0}$ changes significantly due to rotational diffusion against the alignment interaction. This results in a very long self-propulsion direction reversal time which increases exponentially with $\omega_0/D_\theta$. Therefore,  exponential growth of diffusion with $\omega_0$  is expected as long as the orientation-locking barrier is not too large.

 (ii) For large values of $\omega_0$,  an alternative mechanism to exit from the direction-locked state becomes operational. Through this mechanism, thermal translational diffusion against the transverse component of self-propulsion velocity takes the particle away from the region of alignment interaction where they can rotate freely. However, this mechanism needs quite fast rotational diffusion so that the self-propulsion velocity direction can easily be reverted before the particle reaches the opposite wall or returns to the same wall. The direction reversal of $\vec{v_0}$ becomes feasible under the condition $l_\theta \ll y_L$. Therefore, the unlocking and velocity reversal time is expected to be proportional to $\exp\left(\Delta V/D_0 \right)$. Here, the barrier $\Delta V$ is determined by the transverse component of self-propulsion velocity and the cut-off length $\lambda$. Thus, diffusion in the limit $\omega_0 \rightarrow 0$ can be approximated  as,
 \begin{eqnarray}\label{tauR1}
D_{\infty} \sim \frac{v_0^2}{2}  \exp\left(v_0 \lambda/\sqrt{2}D_0 \right)
\end{eqnarray}
This estimation ignores contributions of diffusion when the active particle gets away from the boundaries. However, Eq.~(\ref{tauR1})  well justifies notable features of $D_{\infty}$: exponential growth with $1/D_0$ [see inset of Fig.~5(a)] and $\omega_0$ independence.  %dependence  Our numerical results are well accords with this analytic estimate (indicated by dashed lines in Fig.~5a). Further, the top inset in Fig.~5a depicts $D/D_s$ versus $1/D_0$, which fairly agrees with our analytic prediction Eq.~(\ref{tauR1}).

Diffusion constant as a function of alignment interaction strength for different chiral torques $\Omega$ are presented in Fig.~5(b). When rotational diffusion is not very slow, the key features in $D\; vs. \; \omega_0$ for chiral and achiral particles are alike. They share some common features:  Diffusivity grows noticeably fast as soon as the alignment interaction becomes stronger than $\max{\{D_\theta, \Omega\}}$. Moreover, for very large values of $\omega_0$, diffusion becomes insensitive to changes in $\omega_0$.  However,  it is apparent from the simulation results that chirality drastically enhances diffusion. This is because chiral torques help to realign particles with the stable self-propulsion velocity direction, leading to an increase in the velocity direction reversal and hence diffusion.
\begin{figure}
\centering
\includegraphics[width=0.85\linewidth]{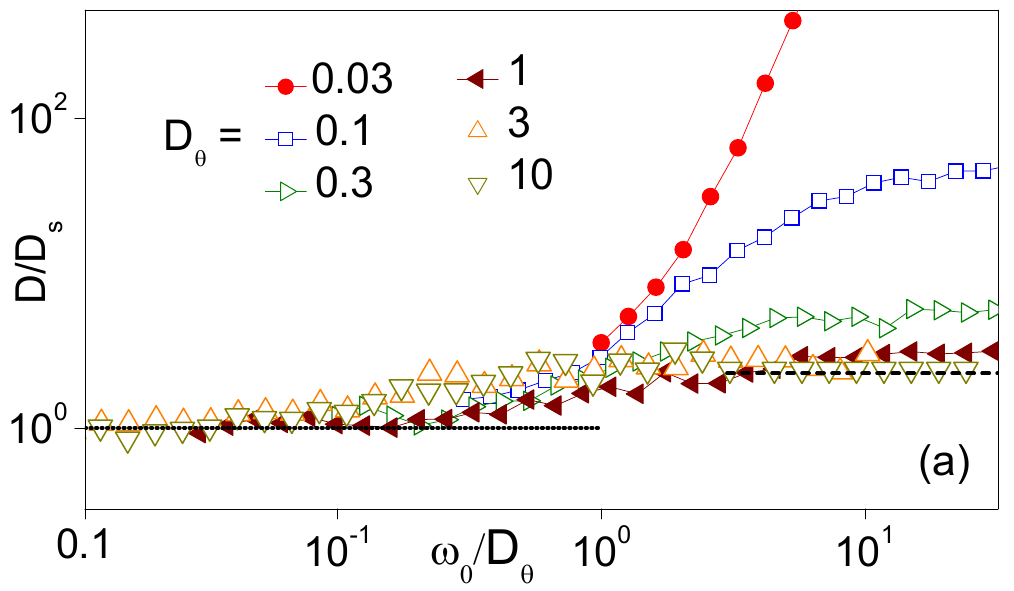}
\includegraphics[width=0.85\linewidth]{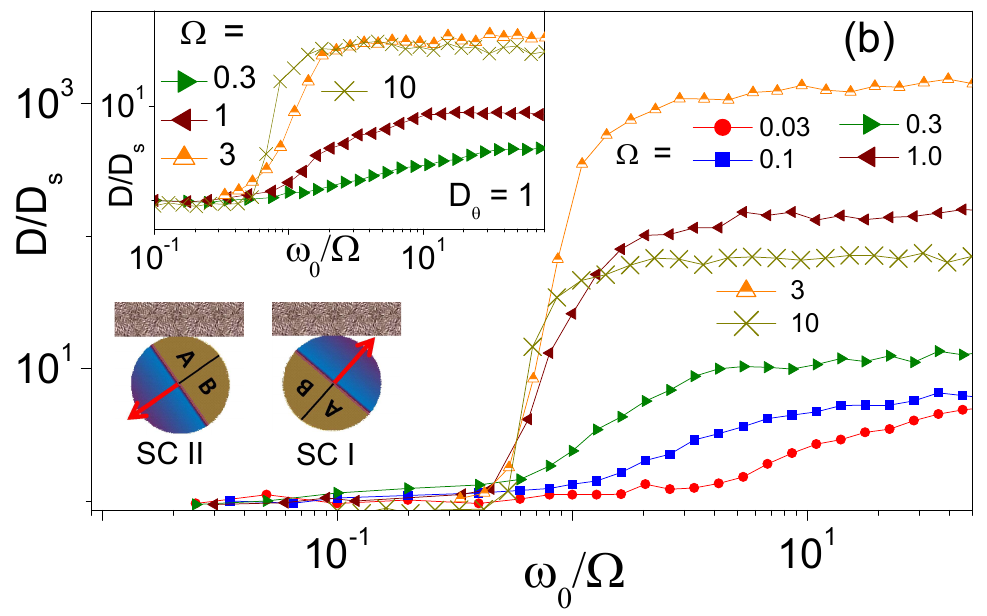}
\caption {(Color online) (a) $D \; vs. \; \omega_0$ of achiral active particles for the Configuration III. Here, dotted  and dashed lines represent analytical estimations based on Eq.~(\ref{free-duffusion}) and (\ref{phi011}), respectively. (b) $D \; vs. \; \omega_0$ for different chiral torque $\Omega$ (see legends) and $D_\theta = 0.3$. The top inset depicts similar plots as the main panel but $D_\theta = 1$. Schematics in the lower inset represent two stable configurations, SC I and SC II, resulting of chiral torques.  Simulation parameters (unless reported otherwise in the legends): $v_0 = 1.0, \; D_0 = 0.01, \; \lambda = 0.05,\; \chi = 0, \;\Omega_0 = 0, \; D_\theta = 0.3, \; {\rm and} \; y_L=1$. }
\end{figure}
\subsection{Configuration III: Stable self-propulsion velocity direction parallel to the boundary walls}
We now examine the scenario where $\vec{v_0}$ is stable against both walls at angles $\theta = 0$ and $\pi$ [as shown in Fig. 2(c)]. As previously mentioned, for this stable structure, the alignment interaction  $V(\theta)$ is modeled by Eq.~(\ref{wall1}), with the degeneracy factor, $\tilde{d}=2$ and $\phi=\pi/2$. The right panel in Fig.~2(c) depicts variation of $V(\theta)$ with $\vec{v_0}$ direction. 
 
Our simulation results, presented in Fig. 6(a), show that when $l_\theta$ is much larger than the channel width $y_L$, the diffusion of achiral active particles start increasing rapidly as soon as the alignment interaction strength exceeds $D_\theta$. On the other hand, when $l_\theta$ is much shorter than $y_L$, the effects of alignment interaction on diffusion get significantly suppressed.

When particles align themselves in their stable configuration near walls, their velocity component perpendicular to the channel axis becomes zero. 
This means that self-propulsion doesn't help particles stay near the wall. As a result, they can easily diffuse out from the alignment interaction region due to thermal translational diffusion. The escape time from this region can be approximated as $\tau_D \sim \lambda^2/3D_0$ \cite{gardiner}. 
When the alignment interaction strength is strong enough, self-propulsion velocity direction reversal can occur only when the particle gets away from the walls. For fast rotational diffusion, where $l_\theta < y_L$, an active particle has enough time to reverse its velocity direction before reaching the opposite wall. Therefore, particles randomly switch between two regions: near the walls where $\vec{v}_0$ is parallel to the channel walls and in the bulk where active particles exhibit correlated random motion.  Considering the independent contribution from these two regions, we can express diffusion as,  
\begin{eqnarray}\label{phi011}
D = p_w v_0^2 \tau_D + p_b \left( \frac{v_0^2}{2D_\theta} + D_0 \right)
\end{eqnarray} 
The first term estimates diffusion contribution within the regions, $y\geq y_L - \lambda$ and $y\leq  \lambda$. This estimation assumes that  $\vec{v_0}$ direction gets reverted as soon as the active particle diffuses away from the region of alignment. The second term in Eq.~(\ref{phi011}) represents the contribution of bulk diffusion. The probability of finding the particle near walls ($y\geq y_L - \lambda$ and $y\leq  \lambda$) and in the bulk ($\lambda \leq y \leq y_L - \lambda $) is denoted by $p_w$ and $p_b$, respectively. They can be approximated as, $p_w \sim 2\lambda/y_L$ and $p_d \sim 1-2\lambda/y_L$ when $D_0 \gg v_0^2/2D_\theta$. Estimation based on the Eq.~(\ref{phi011})for $D_\theta = 100$ [indicated by a dashed line in Fig.~6(a)] well corroborates simulation results.

On the other hand, when rotational diffusion is slow enough, $l_\theta > y_L$, active particles do not get enough time to be reverted while swimming from one side of the channel to the opposite side boundary, where particles get realigned against the wall. In this parameter regime, velocity direction reversal time and hence diffusion rapidly grow with increasing $\omega_0$ and $\tau_\theta$. %As a result, active particle diffusivity grows almost exponentially exponentially.

 Figure 6(b) shows variation of diffusion constant as a function of $\omega_0$ for levogyre active particles for different chiral torques. Here, the main panel (b) and its inset, respectively consider the situations, $l_\theta > y_L/2$ and  $l_\theta \sim y_L/2$. Simulation results show that diffusion  starts sharply enhancing as soon as $\omega_0$ becomes stronger than  $\max{\{\Omega,D_\theta\}}$. For the both limits, $l_\theta > y_L/2$ and  $l_\theta < y_L/2$, diffusion becomes insensitive to the alignment interaction strength when $\omega_0$ is much stronger than rotational diffusion and the chiral torques. However,  $D_\theta$ and $\Omega$ determine asymptotic values of diffusion.

At the stable $\vec{v_0}$ orientation for chiral active particles near walls, the alignment interaction torque is balanced by the chiral torque, resulting in stable self-propulsion velocity directions at the angle,  
\begin{eqnarray}\label{phi01}
\tilde{\theta}=\frac{1}{2}\sin^{-1}\left(\frac{\Omega}{2\omega_0}\right)
\end{eqnarray}
where, $\tilde{\theta}$ accounts for the tilting angles concerning the stable alignment for $\Omega = 0$.  Depending upon the magnitude of $\tilde{\theta}$ and otherwise achiral stable aligned states ( possible ${\vec v_0}$ orientations, $\theta = 0, \; \pi$), the chiral torque can cause the self-propulsion velocity to either help the active particles stay near walls or enable them to move away from walls. The chiral torque results in two stable alignments, which we refer to as SC I and SC II [illustrated in the inset of Fig. 6(b)].

In SC I, self-propulsion pushes particles against the walls, this enhances lifetime of the state. This is attributed to the fact that escaping from SC I requires rotational or translational diffusion against certain barriers. On the other hand, in the SC II, the particle gets pushed away from the alignment interaction zone due to self-propulsion. As a result, SC II becomes short-lived. Therefore, the time it takes for $\vec{v_0}$ to reverse direction, $\tau_d$, can be approximated as the escape time from SC I. Thus, diffusion can be expressed as, $D = v_a^2\tau_d$. Here, $v_a$ is the component of $\vec{v_0}$ along the channel axis, and $\tau_d$ depends on $D_\theta$, $\Omega$, and $D_0$. When $\omega_0 \ll \Omega$, particles assume a stable configuration at angles $\theta = 0$ and $\pi$. In this scenario, $v_a \approx v_0$. Moreover, escaping from these stable configurations is independent of the strength of the alignment interaction. Thus, a plateau is expected (as seen in simulation data) in $D$ versus $\omega_0$. However, the height of the plateau is a non-monotonic function of $\Omega$. %This is because depending upon its amplitude the chiral torque can facilitate escape from SC I, as well as, it forces swimmers to return to the aligned state. 

\section{Summary and concluding remarks}
We explore the potential effects of particle-wall alignment interactions on the diffusion of active particles through a narrow channel. This type of interaction is particularly relevant for artificial active particles, which are composed of two distinct hemispheres made of materials with different dielectric and magnetic properties. Therefore, the strength of the interaction energy is expected to be influenced by the particle's self-propulsion velocity direction with respect to the walls. Further, particle-wall interaction can also occur for natural active particles.

Our study models the particles' dynamics using overdamped Langevin equations and analyzes the diffusion properties for all possible stable orientations of the particles with respect to the walls. Here, we summarize the significant unbiased transport features that emerge due to particle-wall alignment interactions.
\newline (i) For the most stable $\vec{v_0}$ orientation at an angle $\theta =\pi/2$ and $-\pi/2$ near the top and bottom walls, respectively, the particle-walls alignment interactions noticeably suppress diffusion of achiral active particles. We show that for $\omega_0 > D_\theta$, diffusion becomes inversely proportional to the square of $\omega_0$. On the other hand, for chiral active particles, diffusion versus $\omega_0$ passes through a minimum followed by a maximum. However, for $\omega_0 \rightarrow \infty$, diffusion for both chiral and achiral particles becomes equal to the pure thermal bulk diffusion $D_0$.
\newline (ii) When stable self-propulsion velocity directions make acute angles against walls (see Configuration II), diffusion constant as a function of $\omega_0$ first exponentially grows and then reaches a plateau.  The height of the plateau exponentially increases with the inverse of thermal noise strength $1/D_0$. Here, diffusion features of chiral and achiral particles are alike when alignment interaction is much stronger than the chiral torque. However, for chiral active particles a sudden jump is observed in $D \; vs. \; \omega_0$ as soon as the aligned interaction strength gets stronger than the chiral torque.     
\newline (iii) For the stable orientation near the boundary wall corresponding to the configuration III ($\vec{v_0}$ is parallel or anti-parallel to the channel axis), diffusion behavior is noticeably different for $l_\theta \ll y_L$ from its opposite limit. When $l_\theta \gg y_L$ self-propulsion velocity reversal time and hence diffusion grow very fast with the alignment interaction strength. However, in the opposite limit where $y_L$ is much larger than $l_\theta$, diffusion first slowly grows with $\omega_0$ and finally reaches an asymptotic value.
\newline (iv) When the most stable $\vec{v_0}$ orientations near the walls such that it takes particles away from the alignment interaction zone, diffusion features of the active particles are not affected due to the alignment interaction (results are not shown). 

Our analysis focuses on the dilute suspension of active particles where inter-particle interactions can safely be ignored. However, in a densed suspension, inter-particle interaction could impact the transport feature by assisting motility-induced phase separation(MIPS). Previous studies \cite{pbag,Debnathnano} show that confinement induces MIPS even without particle-wall alignment interactions. As the alignment interaction can detain particles around the boundary for a long time, we expect notable effects on collective behaviors when dense suspension of active particles is placed in narrow channels.    

%\section{Conclusions}
In conclusion, we systematically demonstrate the possible impacts of particle-wall alignment interactions on unbiased transport through narrow channels. Our findings, which are summarized above, would help us better understand the transport control mechanism of active particles through narrow structures aiming at targeted drug delivery and other cutting-edge nano-technological applications\cite{Mishra,Bunea,Dabbagh}. Further, our study brings up several related issues for future work. Specifically, can particle-wall interaction break spatial symmetry to induce autonomous directed motion or other non-equilibrium phenomena? How can particle-wall interaction be exploited to facilitate the transport of active particles? Another pertaining issue is the collective motion of active particles through narrow channels. What would be the impacts of particle-wall alignment interaction in MIPS and the flocking of active particles through narrow channels? These issues are crucial for gaining a deeper understanding of transport mechanisms in narrow confined structures.

% \section*{Author contributions}
% We strongly encourage authors to include author contributions and recommend using \href{https://casrai.org/credit/}{CRediT} for standardised contribution descriptions. Please refer to our general \href{https://www.rsc.org/journals-books-databases/journal-authors-reviewers/author-responsibilities/}{author guidelines} for more information about authorship.
\section*{Author Contributions}
Poulami Bag: Conceptualization (equal); Data curation (equal);
Formal analysis (equal);  Investigation
(equal); Methodology (equal);  Writing -- review and editing (equal). Shubhadip Nayak:
Data curation (equal); Formal analysis (equal); Investigation (equal). Pulak K. Ghosh: Conceptualization (equal); Formal analysis (equal); Investigation (equal); Supervision (equal); Writing -- original draft (equal); Writing -- review and editing (equal).

\section*{Data availability}
The data that support the findings of this study are available within the article.

\section*{Conflicts of interest}
The authors have no conflicts to disclose.

\section*{Acknowledgements}
P.K.G. is supported by SERB Core
Research Grant No. CRG/2021/007394. P.B. thanks UGC, New Delhi, India, for the award of a Junior Research Fellowship. 

%%%END OF MAIN TEXT%%%

%The \balance command can be used to balance the columns on the final page if desired. It should be placed anywhere within the first column of the last page.

\balance

%%%REFERENCES%%%
\bibliography{rsc}  

\begin{mcitethebibliography}{83}
\bibitem[not()]{note2}
Although $\omega_0$ has a dimension of angular velocity, in the overdamped limit, this quantity is the measure of alignment interaction torque as well as alignment interaction strength. Also, $\Omega$ is assumed to be equivalent to chiral torque.\relax
\mciteBstWouldAddEndPunctfalse
\mciteSetBstMidEndSepPunct{\mcitedefaultmidpunct}
{}{\mcitedefaultseppunct}\relax
\EndOfBibitem
\end{mcitethebibliography}
\bibliographystyle{rsc} %the RSC's .bst file

\end{document}